\documentclass[useAMS,usenatbib,usegraphicx]{mn2e_modmargin}

\usepackage{lscape}
\usepackage{color}
\usepackage{subfigure}
\usepackage{amssymb}
\usepackage{amsmath}
\usepackage{url}
\usepackage{amsfonts}
\usepackage{amsbsy}
\usepackage{graphicx}
\usepackage{subfigure}
\usepackage{verbatim}
\usepackage{multicol}

\title[PAHs in Seyfert 2 NGC\,1808]{Polycyclic Aromatic Hydrocarbon in the Central Region of the Seyfert 2 Galaxy NGC\,1808}
\author[Dinalva A. Sales et al.]{Dinalva A. Sales$^{1,2}$\thanks{dassps@rit.edu};M. G. Pastoriza$^{1,3}$
; R. Riffel$^{1}$ and Cl\'audia Winge$^{4}$\\
$^{1}$Departamento de Astronomia, Universidade Federal do Rio Grande do Sul. Av. Bento Gon\c calves 9500, Porto Alegre, RS, Brazil\\
$^{2}$Department of Physics, Rochester Institute of Technology, 84 Lomb Memorial Drive, Rochester, NY 14623, USA\\
$^{3}$Conselho Nacional de Desenvolvimento Cient\' ifico e Tecnol\' ogico, Brazil\\
$^{4}$Gemini Observatory, c/o Aura, Inc., Casilla 603, La Serena, Chile}

\begin{document}

\date{Accepted 1988 December 15. Received 1988 December 14; in original form 1988 October 11}

\pagerange{\pageref{firstpage}--\pageref{lastpage}} \pubyear{2012}

\maketitle

\label{firstpage}

\begin{abstract}

We present mid infrared (MIR) spectra of the Seyfert 2 (Sy~2) galaxy NGC\,1808, obtained with the 
Gemini's Thermal-Region Camera Spectrograph (T-ReCS) at a spatial resolution of $\sim$26\,pc.
The high spatial resolution allowed us to detect bright polycyclic aromatic hydrocarbons (PAHs)
emissions at 8.6$\mu$m and 11.3$\mu$m in the galaxy centre ($\sim$26\,pc) up to a radius of
70\,pc from the nucleus. The spectra also present [Ne\,{\sc\,ii}]12.8$\mu$m 
ionic lines, and H$_{2}\,S$\,(2)12.27$\mu$m molecular gas line. We found 
that the PAHs profiles are similar to Peeters's $A$ class, with the line peak shifted
towards the blue. The differences in the PAH line profiles also suggests that the molecules 
in the region located 26\,pc NE of the nucleus are more in the neutral than in the 
ionised state, while at 26\,pc SW of the nucleus, the molecules are mainly in ionised state. 
After removal of the underlying galaxy contribution, the nuclear spectrum can be represented 
by a Nenkova's clumpy torus model, indicating that the nucleus of NGC\,1808 hosts 
a dusty toroidal structure with an angular cloud distribution of $\sigma = 70^{\circ}$, 
observer's view angle $i = 90^{\circ}$, and an outer radius of R$_{0}\sim$0.55\,pc. 
The derived column density along the line of sight is 
N$_H$\,=\,1.5\,$\times\,10^{24}$\,cm$^{−2}$, which is sufficient to block the hard radiation from the active nucleus, and would explain 
the presence of PAH molecules near to the NGC\,1808's active nucleus.
\end{abstract}


\section{Introduction}

Mid-infrared (MIR) spectra of active galactic nuclei (AGN) 
are dominated by the emission features at 3.3, 6.2, 7.7, 8.6, 11.3 and 12.7\,$\mu$m,
commonly called the unidentified infrared (UIR) bands \citep[e.g.][]{gillett73,geballe85,cohen86}, and generally attributed to Polycyclic Aromatic Hydrocarbon (PAHs) molecules 
\citep{leger84,puget89,allamandola89,allamandola99}. The spectra of AGN also present prominent molecular hydrogen 
and forbidden emission lines \citep[]{sturm00,weedman05,wu09,gallimore10,sales10}, as well as silicate bands at $\sim$ 9.7\,$\mu$m and 18\,$\mu$m, both in emission and/or absorption. 

The absence of PAH emission close to the central engine of active galaxies has been attributed to the destructive 
effects of highly energetic UV photons (h$\nu\,\sim\,$40\,eV) supplied
by the active nucleus and, therefore, the presence of these features have been mostly used as a tracer of massive star formation \citep[e.g.][]{peeters04,siebenmorgen04,allamandola99,tielens08}.
However, \citet{avoit92,bvoit92} has shown that PAHs can survive if the hard AGN radiation is shielded by dusty gas with high column density. Recently, \citet{sales10} showed that 
a large fraction ($>$\,70\%) of Seyfert galaxies shows bright PAH emission lines, and demonstrated that PAHs in AGN present higher ionisation fraction and larger molecules ($>$ 180 
carbon atoms) than in Starburst galaxies.

The unified model of AGNs proposes the existence of a dense concentration of 
absorbing material surrounding the central engine in a toroidal distribution, 
which blocks the broad line region (BLR) from the line of sight in 
Seyfert 2 (Sy\,2) galaxies \citep[see][]{antonucci93}. Nowadays such toroidal 
structure is believed to be composed by gas clumps or clouds \citep[e.g.][]{honig06,nenkova02}. In fact, very recently we  \citep{sales11} have shown that the nuclear spectral energy distribution of NGC\,3281, a Sy~2 galaxy, can be well described by clumpy torus models with about 14 dusty clouds in the torus equatorial radius. Similar results are also obtained for other 
AGN \citep[e.g.][]{ramos09,ramos11,nikutta09}.



It has been known by approximately five decades that NGC 1808 presents a peculiar nuclei, evidenced by the detection of star forming regions disposed in a circumnuclear ring, the so called {\it hot spots} \citep{pastoriza65,pastoriza67a,pastoriza67b}.
In addition, \citet{galliano05} detected bright MIR sources in the central region of NGC\,1808, associated with already known radio sources \citep{saikia90}, suggesting that these sources are young embedded star clusters. The presence of
an active nucleus in this source is supported by the presence of extended wings on both H$_{\alpha}$
and [N{\sc\,ii}] emission lines \citep{veron85}.
These authors found that NGC\,1808 falls in the Seyfert region of the 
\citet{baldwing81} diagram with H$\alpha$/[N{\sc ii}]$\lambda\,6583=0.94\,pm\,0.05$

and presents the full width at half maximum (FWHM) of H$\alpha$ component approximately 550\,km/s.
In addition, \citet{awaki96} have detected hard X-ray emission variability suggesting 
the presence of a low-luminosity AGN in NGC\,1808.
This object is a nearby galaxy (d=13\,Mpc, using H$\rm _{0}=74 km\,s^{-1}\,Mpc^{-1}$ ) exhibiting evidence that strong star formation and an active nucleus coexist in the central regions,  making it a key object to investigate PAH emission and torus structure through N band spectroscopy. 

Here we present a detailed study of N band spectra of NGC\,1808 using ground-based MIR high angular resolution, R$\sim100$ spectra obtained with the Thermal-Region Camera Spectrograph \citep[T-ReCS;][]{telesco98} at the 8.0m Gemini South telescope, achieving a spatial resolution of 0\farcs36, or $\sim26$pc at NGC\.1808 distance, which makes our data quite adequate to study the molecular and dust distribution in the inner $\sim$\,100\,pc  of this galaxy. This paper is structured as follows: in Section \ref{observation} we describe the observation and data reduction processes. Results are discussed in Section \ref{results}. Summary and conclusion are presented in Section \ref{conclusions}.

\section{Observations and Data Reduction}\label{observation}

\begin{figure}
\centering
\includegraphics[width=6cm]{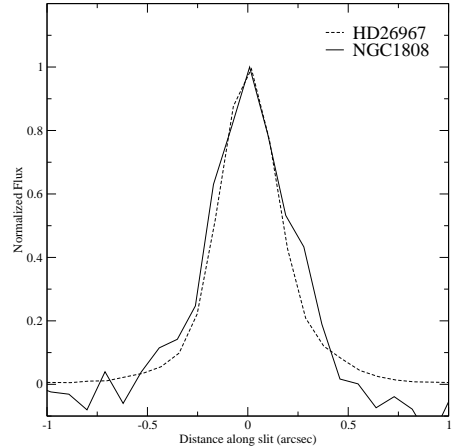}
\caption{The $N$-band spatial emission profile (solid line) of NGC\,1808 compared with that of the telluric star
HD\,26967 (dotted line). The fluxes were normalised to the peak value.}
\label{profile}
\end{figure}

The $N$-band spectra of NGC\,1808 were obtained with the T-ReCS in queue 
mode at Gemini South, in 2009 August 28 and 29 UT, as part of program GS-2009B-Q-19. 
The observational conditions were photometric, with water vapour column in the range 5--8mm, and image 
quality of 0\farcs34 in the $N$-band, measured from the 
telluric standard acquisition images observed before NGC\,1808. 
The $N$-band luminosity profile of the galaxy is slightly extended with 
respect to that of HD\,26967 telluric star (see Fig. \ref{profile})

We used a standard chop-nod technique to remove time-variable 
sky background, telescope thermal emission, and the effect of 1/\textit{f} noise 
from the array/electronics. In order to include 
only the signal of the guided beam position in the frame and avoid possible nod 
effects in the spatial direction the slit were oriented along P.A.\,=\,45\degr with 
a chop throw of 15\arcsec, oriented along P.A.\,=\,45\degr. The same slit 
position/nod orientations was used for the telluric standards. 
The T-ReCS configuration used was the low-resolution 
($R\,\sim$\,100) grating and the 0\farcs36 slit, for a dispersion of 0.0223\,$\mu$m 
pixel$^{-1}$ and a spectral resolution of 0.08\,$\mu$m. The pixel scale is 0.089\arcsec 
pixel$^{-1}$ in the spatial direction, and the slit is 21\farcs6 long. The resulting 
spectral coverage is between 8-13$\mu$m centred at 10.5$\mu$m. The total on-source 
integration time was 900s.

We used the {\sc midir} and {\sc gnirs} sub-packages 
of Gemini IRAF\footnote{IRAF is distributed by the National Optical Astronomy 
Observatory, which is operated by the Association of Universities for Research 
in Astronomy (AURA), Inc., under cooperative agreement with the National Science
Foundation.} package to reduce the data. In addition, to extract the final 
spectrum we combined the chop- and nod-subtracted frames using the tasks {\sc tprepare} 
and {\sc mistack}. Wavelength calibration was obtained from the skylines in 
the raw frames. To remove the telluric absorption lines from the galaxy spectrum, 
we divided the extracted spectrum for each observing night by that of the telluric 
standard star HD\,26967 \citep{cohen99}, observed before the science 
target. The flux calibration used the task {\sc mstelluric} that 
interpolates a blackbody function to the spectrum of the telluric standard in order 
to derive the proper instrumental response.

\begin{figure*}
\centering
\begin{tabular}{cc}
\multicolumn{2}{c}{\includegraphics[width=6cm]{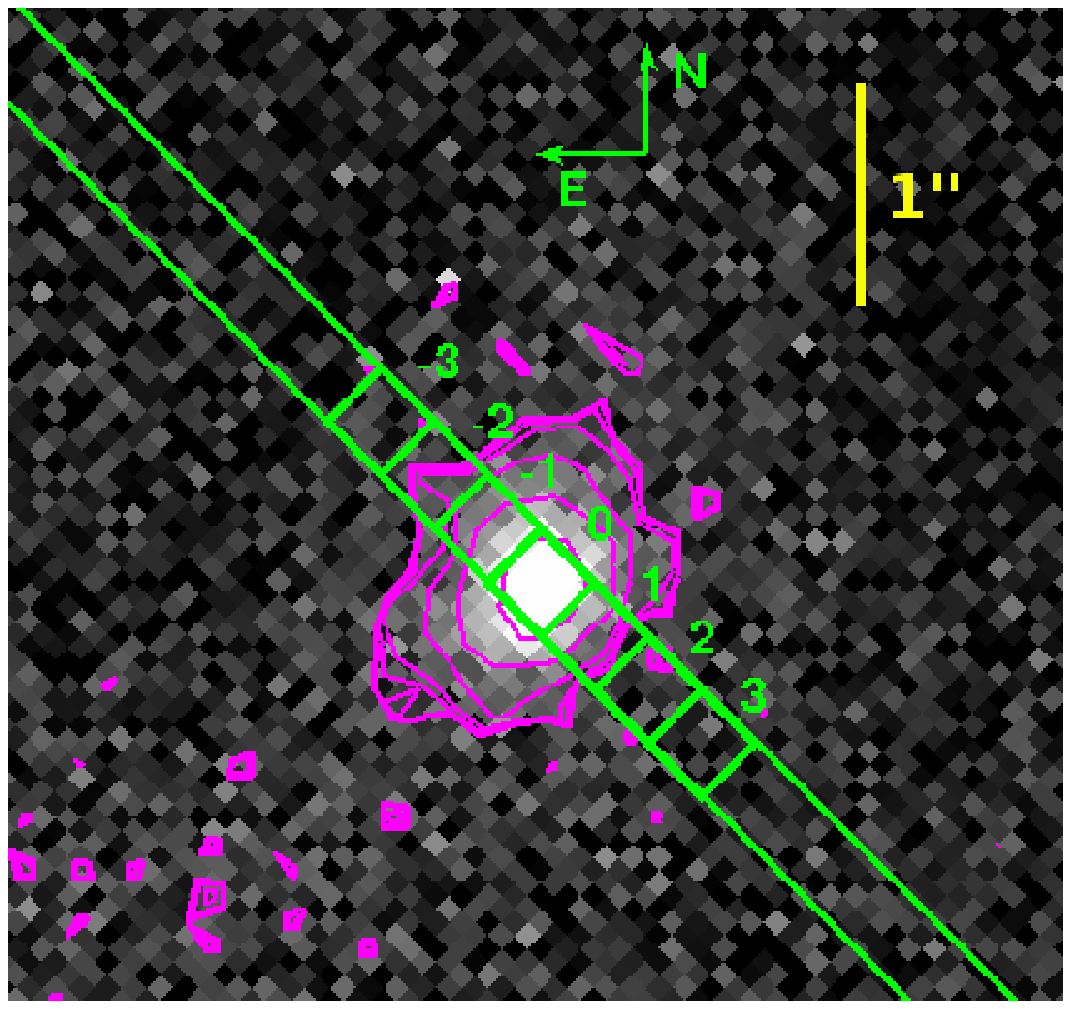}}\\
\multicolumn{2}{c}{\includegraphics[width=15cm]{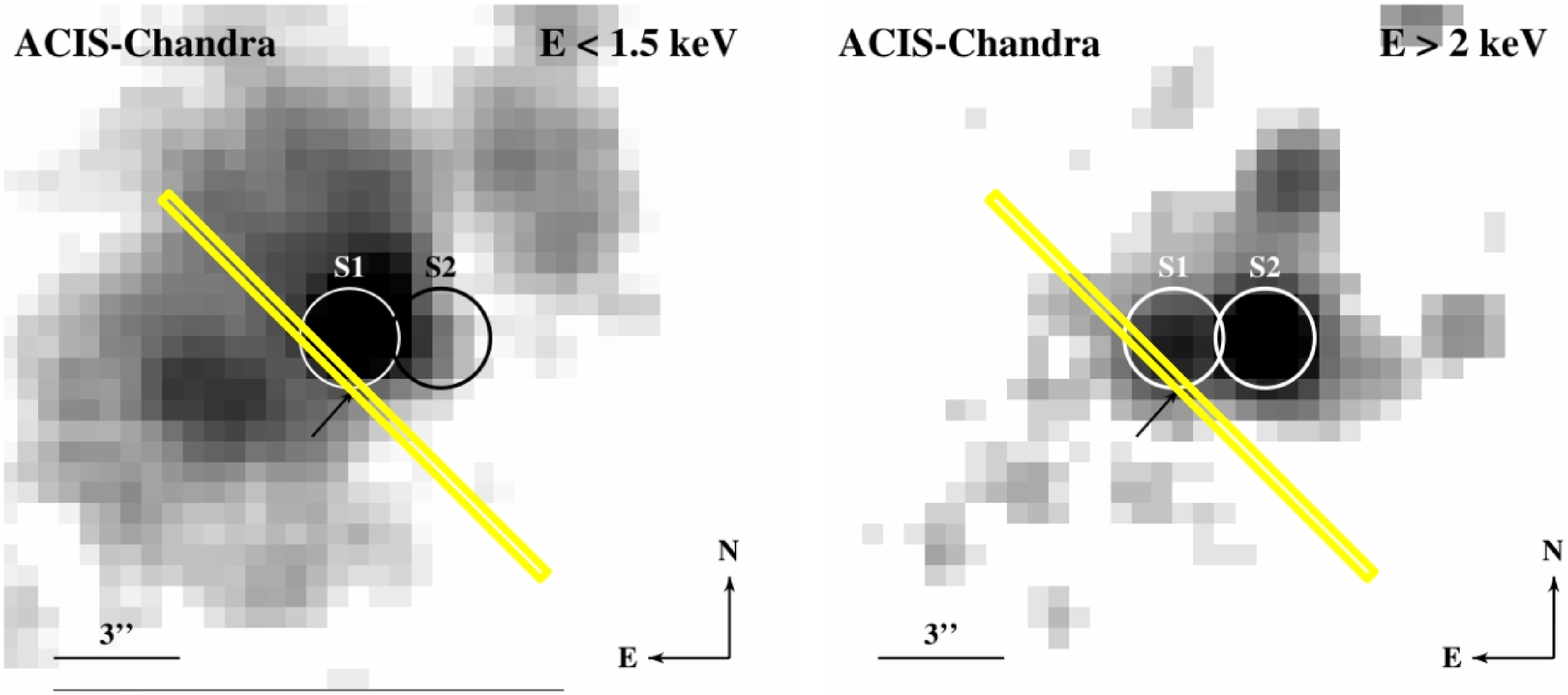}}
\end{tabular} 
\caption{The top panels shows the T-ReCS long-slit position over-plotted on the contoured 
NGC\,1808 acquisition image. The contours are linear and stepped at 20\% of the peak.
The bottom panels show the T-ReCS slit superposed to the Chandra ACIS soft (left) and hard (right) X-ray 
images from \citet{jimenez05}.}
\label{fenda}
\end{figure*}

In Fig. \ref{fenda} we show the T-ReCS slit position superimposed on our N band 
acquisition image, as well as the hard ($>$ 2keV) and soft ($<$ 1.5keV) X-ray
images from \citet{jimenez05}. We can see that the X-ray images show two point-like emission 
sources offset from the NGC\,1808 nucleus (see in Figure~\ref{fenda}). These emitting regions were also detected 
in the radio and infrared images \citep{saikia90,kotilainen96} and have been attributed to 
supernova remnants (S1) and H{\sc ii} regions (S2) \citep{saikia90,kotilainen96,jimenez05}.
The T-ReCS slit position crosses the soft X-ray source S1 and diffuse X-ray emission.

\section{Results and Discussion}\label{results}
\subsection{Emission Lines Measurements and Radial Gradients}\label{decomposition}

\begin{figure*}
\centering
\includegraphics[scale=0.62,angle=-90]{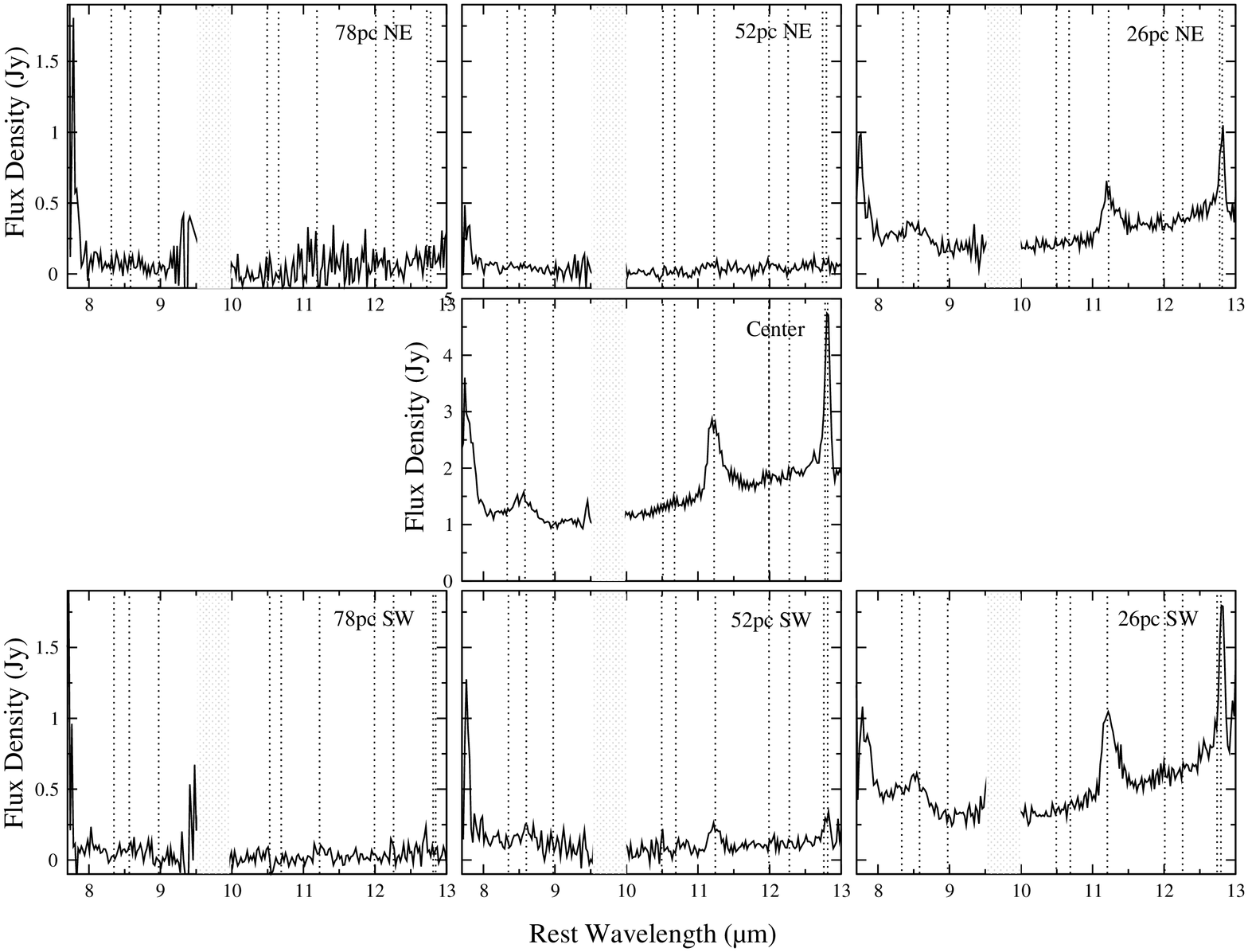}
\caption{Spectra of NGC\,1808 extracted in 26\,pc steps along P.A.\,=\,45$^{\circ}$. 
The dashed lines indicate the positions of the PAH bands, ionic as well as hydrogen molecular lines
that are listed at Tabs.~\ref{lines1} and \ref{lines2}. The telluric O$_{3}$ band is 
represented by the hatched area.}
\label{ngc1808}
\end{figure*}

We extracted 7 one dimensional spectra spaced four pixels or 0\farcs36 ($\sim$\,26 pc) along the slit: one centred in the unresolved nucleus, and six extractions centred at 26\,pc, 52\,pc, and 78\,pc to the northeast (NE) and southwest (SW) directions. 
The spectra are presented in Fig. \ref{ngc1808}, and clearly show intense emission in the PAH bands (8.6\,$\mu$m, 11.3\,$\mu$m and 12.7\,$\mu$m), [Ne{\sc\,ii]}\,12.8\,$\mu$m ionic line, as well as the molecular hydrogen rotational line, H$_{2}\,S$\,(2) at 12.27\,$\mu$m, in the nuclear extraction, and at 26\,pc 
in both NE and SW directions. Weak PAH bands are also 
observed at 52\,pc SW direction of NGC\,1808, while the outer extractions at 78\,pc NE and SW
do not show any emission lines. This result indicates that in this galaxy the PAH bands are emitted
inside a region less than 70\,pc  from the active nucleus, and clearly show one of the most rich 
PAH features at 8.6, 11.3 and 12.7\,$\mu$m observed using T-ReCS so far in AGNs. 

To disentangle the different spectral components in each 1D spectra, both emission 
lines and continuum, and determine the contribution of each component to the spectral 
energy  distribution (SED) of NGC\,1808 as a function of the distance to the galaxy 
centre, we used the {\sc pahfit}\footnote{Source code and documentation for {\sc pahfit} are 
available at http://tir.astro.utoledo.edu/jdsmith/research/pahfit.php} code \citep{smith07} routines.
The input flux uncertainties required by {\sc pahfit} were
assumed to be 10\% of the flux observed, which are the expected ones for T-ReCS 
observations  \citep[see][]{radomski02,mason06}.

\begin{figure*}[htb]
 \begin{centering}
  \begin{tabular}{cc}
   \subfigure[ ]{\includegraphics[clip=true,width=8cm]{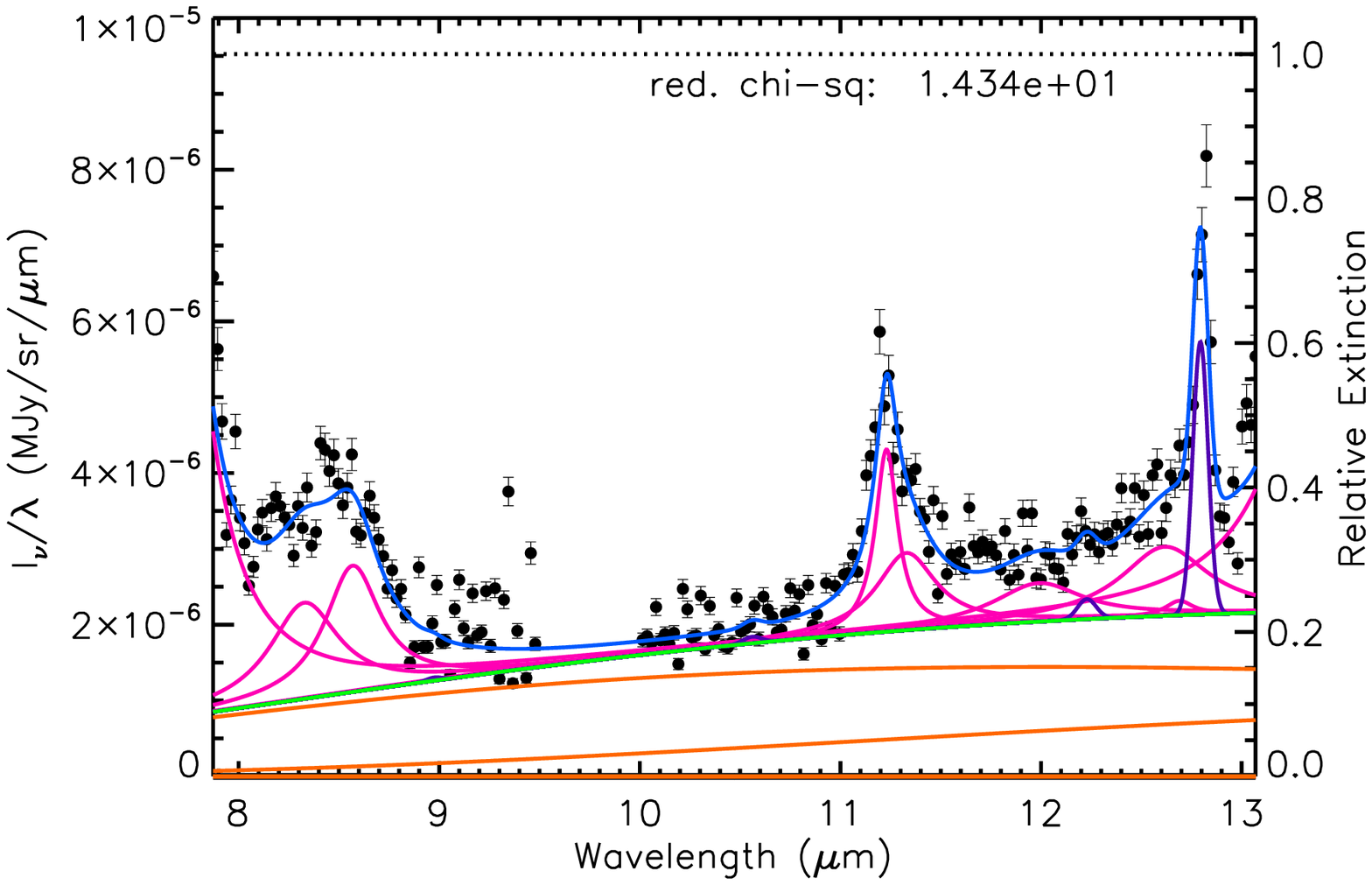}}&
    \subfigure[ ]{\includegraphics[clip=true,width=8cm]{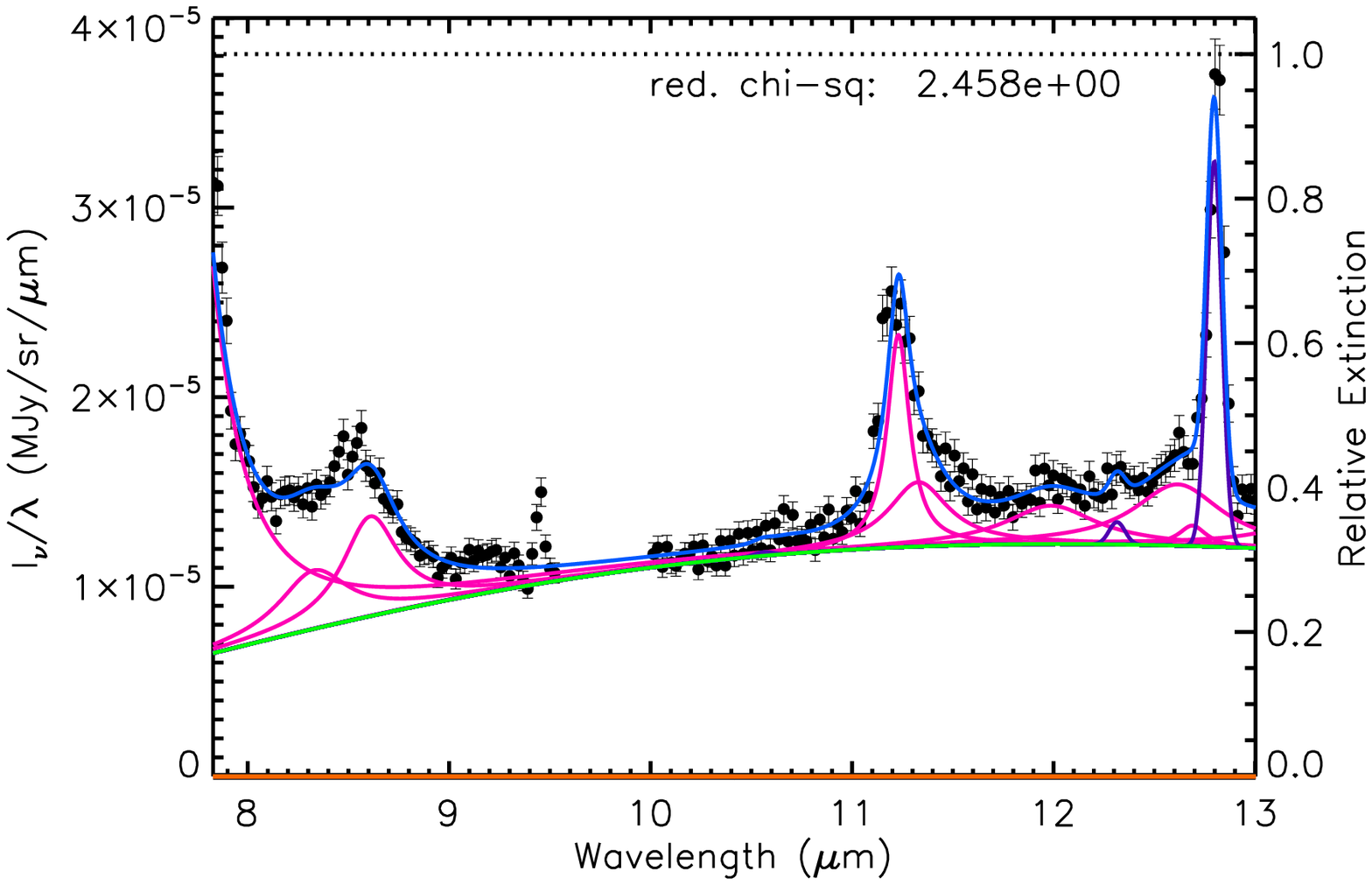}}\\
   \subfigure[ ]{\includegraphics[clip=true,width=8cm]{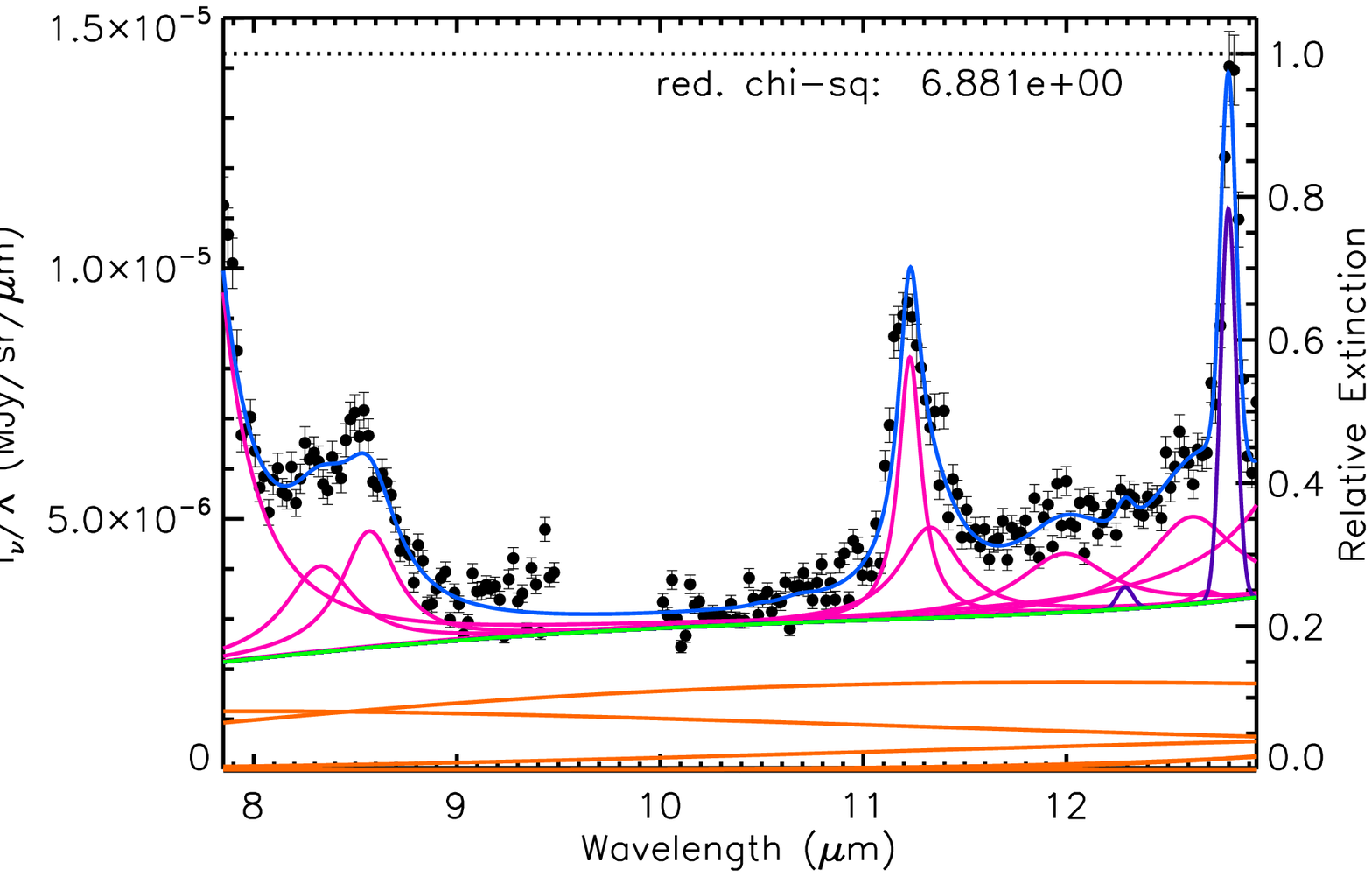}}&
    \subfigure[ ]{\includegraphics[clip=true,width=8cm]{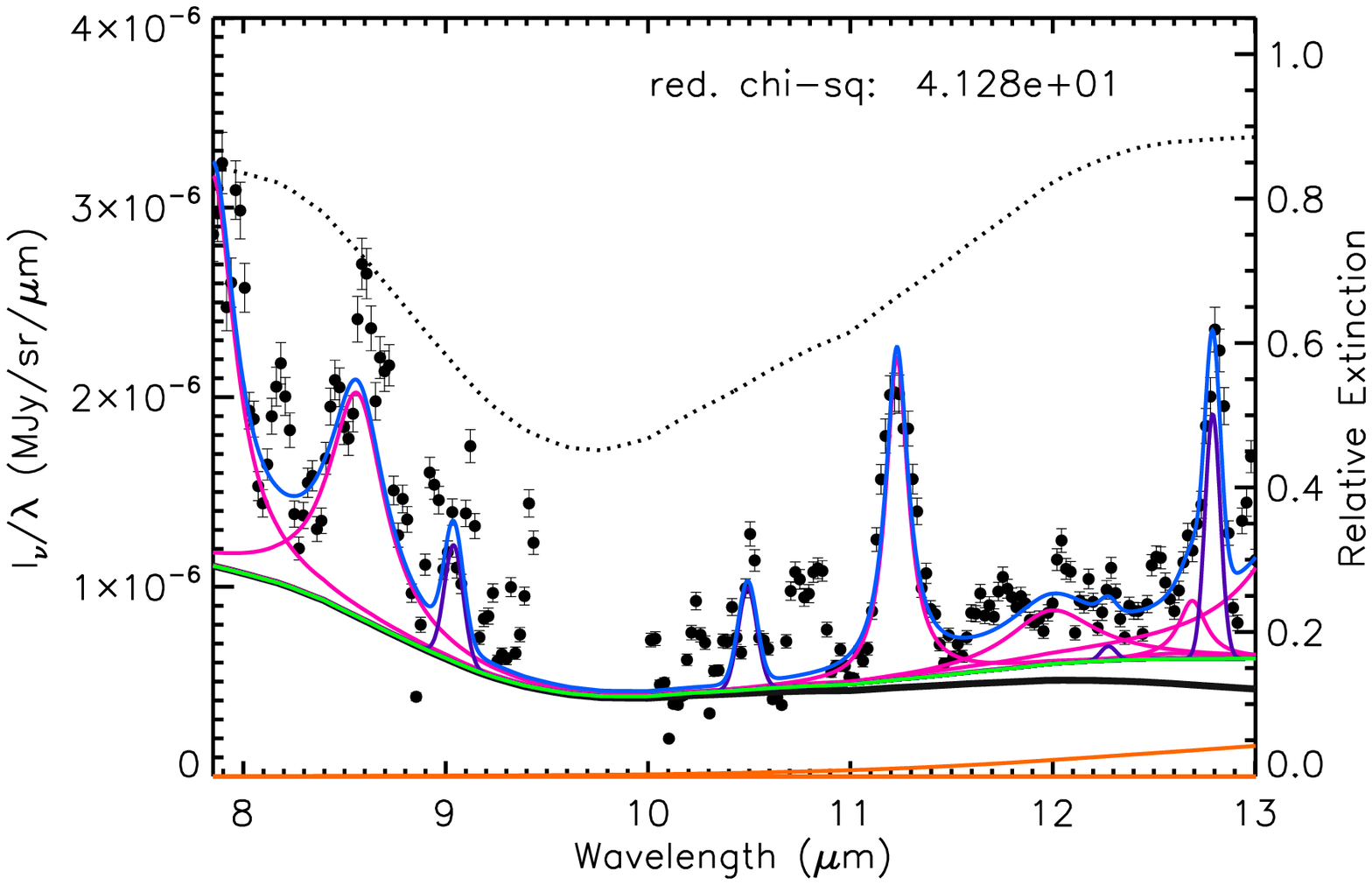}}\\
   \end{tabular}
   \par
  \end{centering}\vspace{-0.3cm}
\caption{Result of the {\sc pahfit} decomposition of the NGC\,1808 spectra: (a) 26\,pc SW; (b) unresolved nucleus; (c) 26\,pc NE: and (d) and 52\,pc NE extractions, respectively. The data points are represented by the dots, with uncertainties plotted as vertical error-bars. The best fit model is represented by the blue line. The dotted black line indicates the mixed extinction components, while solid green and orange lines represent total and individual thermal dust continuum components, respectively. The violet and magenta lines represent the ionic and PAH emission 
lines. Note that {\sc pahfit} code uses the violet solid line to represent the H$_2$ emission as well.}
\label{spectra_pahfit}
\end{figure*}

The {\sc pahfit} code assumes that the spectrum is formed by continuum emission from 
dust and starlight, prominent emission lines, individual and blended PAH emission 
bands, and that the light is attenuated by extinction due to silicate grains. 
The code use the dust opacity law of \citet{kemper04}, 
where the infrared extinction is considered as a power law plus silicate features 
peaking at 9.7\,$\mu$m. For more details see \citet{smith07}.
The remaining input parameters are the same as those used by \citet{sales10}, which 
are appropriate for AGNs. The results of the decomposition of the NGC\,1808 spectra 
are shown in Fig.~\ref{spectra_pahfit}, and the derived emission line fluxes are given 
in Tabs. \ref{lines1} and \ref{lines2}. It is worth note that 10.7$\mu$m PAH emission at the 
centre and 26\,pc NE is at the detection limits and its emission profile can not be
seen in the PAHFIT fitting at Fig.~\ref{spectra_pahfit}. In addition, note that the 11.3$\mu$m PAH emission complex is composed 
by 11.2$\mu$m and 11.3$\mu$m emission bands, which can be seen in Fig.~\ref{spectra_pahfit}.

\begin{table*}
\renewcommand{\tabcolsep}{1.4mm}
\scriptsize
\caption{PAH Emission Line Strengths ($10^{-13}\,W\,m^{-2}$) and Equivalent Widths ($\mu m$).\label{lines1}}
\begin{minipage}[b]{1.0\linewidth}
\begin{tabular}{lccccccccccccccccccc}
\noalign{\smallskip}
\hline\hline
Position & Label & \multicolumn{2}{c}{\underline{~~~~~~~~8.3$\mu$m~~~~~~~~}} & \multicolumn{2}{c}{\underline{~~~~~~~~8.6$\mu$m~~~~~~~~}} & \multicolumn{2}{c}{\underline{~~~~~~~~10.7$\mu$m~~~~~~~~}} & \multicolumn{2}{c}{\underline{~~~~~~~~11.3$\mu$m\footnotemark[1]~~~~~~~~}} & \multicolumn{2}{c}{\underline{~~~~~~~~12.0$\mu$m~~~~~~~~}} & \multicolumn{2}{c}{\underline{~~~~~~~~12.7$\mu$m~~~~~~~~}}\\
&&Flux&EW&Flux&EW&Flux&EW&Flux&EW&Flux&EW&Flux&EW\\
\hline
26\,pc SW & -1 & 3.25\,$\pm$\,0.1 & 0.562 & 3.15\,$\pm$\,0.1 & 0.517 & -		   & -     & 2.95\,$\pm$\,0.2 & 0.396 & 1.19\,$\pm$\,0.1 & 0.163 & 2.05\,$\pm$\,0.3 & 0.291\\
Center    & 0 & 5.35\,$\pm$\,0.8 & 0.119 & 11.2\,$\pm$\,0.7 & 0.242 & 0.29\,$\pm$\,0.3 & 0.006 & 11.8\,$\pm$\,1.2 & 0.251 & 4.85\,$\pm$\,0.7 & 0.110 & 7.54\,$\pm$\,1.6 & 0.182\\
26\,pc NE & 1 & 4.33\,$\pm$\,0.3 & 0.328 & 4.44\,$\pm$\,0.2 & 0.334 & 0.11\,$\pm$\,0.1 & 0.009 & 5.70\,$\pm$\,0.4 & 0.477 & 2.46\,$\pm$\,0.5 & 0.208 & 3.46\,$\pm$\,1.1 & 0.285\\
52\,pc NE & 2 &-		     & -     & 3.13\,$\pm$\,0.1 & 0.990 & -		   & -     & 1.32\,$\pm$\,0.1 & 0.775 & 0.81\,$\pm$\,0.1 & 0.493 & 0.39\,$\pm$\,0.1 & 0.265\\
\hline
\end{tabular}
\\ \footnotemark[1]{The 11.3$\mu$m PAH emission complex is composed by 11.2$\mu$m and 11.3$\mu$m emission bands.}
\end{minipage}
\caption{Ionic and Molecular Emission Line Strengths ($10^{-14}\,W\,m^{-2}$) and Equivalent Widths ($\mu$m).\label{lines2}}
\begin{minipage}[b]{1.0\linewidth}
\begin{tabular}{lccccccccc}
\noalign{\smallskip}
\hline\hline
Position & Label &\multicolumn{2}{c}{\underline{~~~~~~~~[Ar{\sc\,iii]}$8.99\mu$m~~~~~~~~}} & \multicolumn{2}{c}{\underline{~~~~~~~~[S{\sc\,iv]}$10.5\mu$m~~~~~~~~}} & \multicolumn{2}{c}{\underline{~~~~~~~~H$_{2}S$(2)$12.27\mu$m~~~~~~~}} & \multicolumn{2}{c}{\underline{~~~~~~~~[Ne{\sc\,ii]}$12.8\mu$m~~~~~~~}}\\
&&Flux&EW&Flux&EW&Flux&EW&Flux&EW\\
\noalign{\smallskip}
\hline
\noalign{\smallskip}
26\,pc SW & -1 & 0.20\,$\pm$\,0.10 & 0.004 & 0.28\,$\pm$\,0.19 & 0.005 & 0.79\,$\pm$\,0.25 & 0.016 & 8.08\,$\pm$\,0.46 & 0.170 \\
Center    & 0  & -		      & -     & 0.64\,$\pm$\,0.10 & 0.001 & 3.13\,$\pm$\,1.40 & 0.011 & 45.6\,$\pm$\,2.19 & 0.166 \\
26\,pc NE & 1  & -		      & -     & 0.28\,$\pm$\,0.11 & 0.001 & 0.98\,$\pm$\,0.46 & 0.012 & 17.5\,$\pm$\,9.17 & 0.210 \\
52\,pc NE & 2  & 3.88\,$\pm$\,0.18 & 0.235 & 2.80\,$\pm$\,0.13 & 0.238 & 0.24\,$\pm$\,0.08 & 0.022 & 3.18\,$\pm$\,0.16 & 0.326 \\
\hline
\end{tabular}
\end{minipage}
\end{table*}

\begin{figure*}
\centering
\includegraphics[scale=0.4]{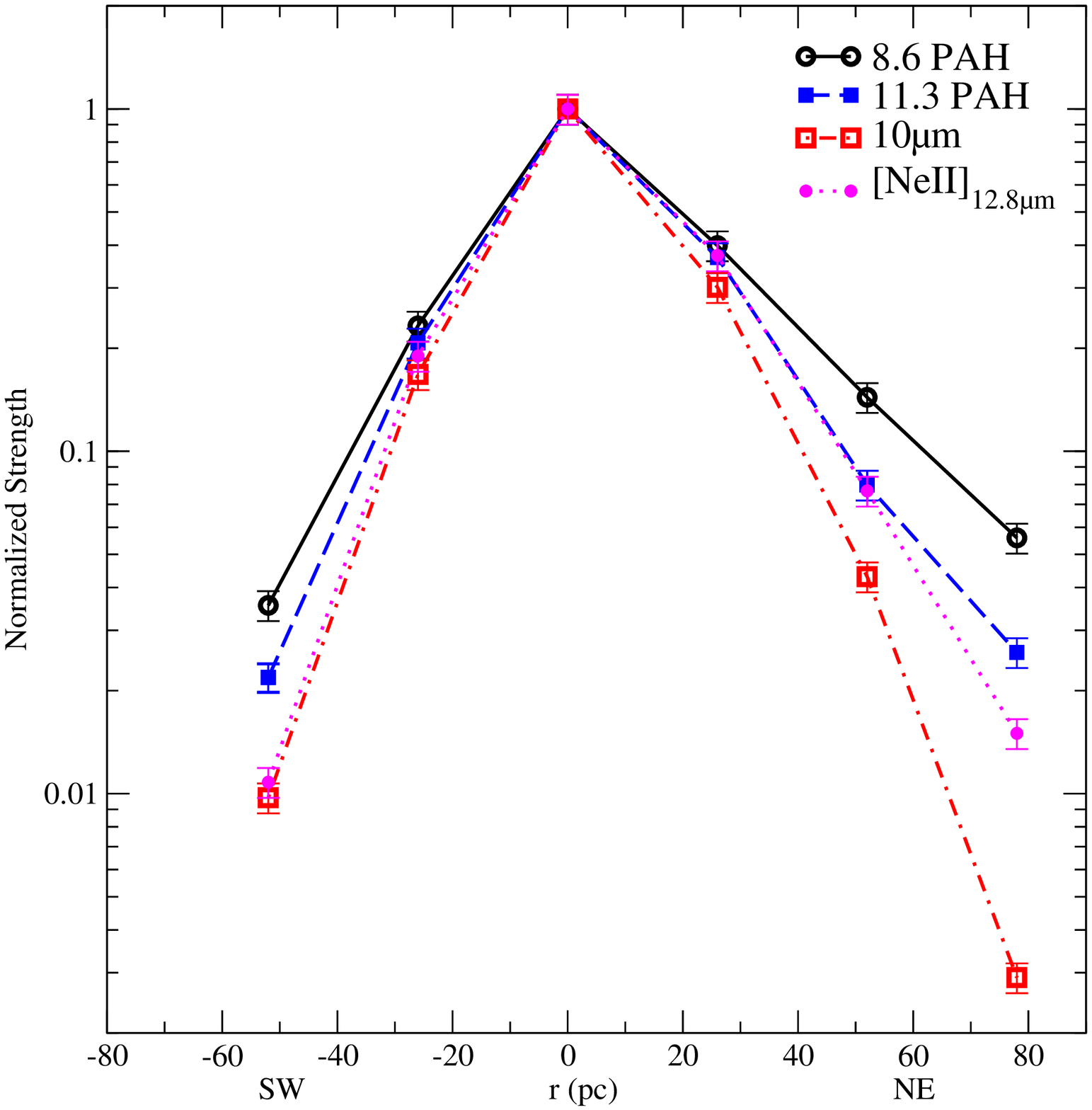}
\caption{Emission lines and 10$\mu$m continuum luminosity profiles along the slit.}
\label{line_profile}
\end{figure*}

Fig. \ref{line_profile} shows the PAH bands and [Ne{\sc\,ii]}\,12.8\,$\mu$m emission line integrated fluxes, as well as the  
10$\mu$m flux continuum emission flux,  as function of distance to the nucleus. Both emission lines and continuum present a strong decreasing gradient from the centre to the outer regions. The similarity between the luminosity profile of the PAH emission bands, ionic line and continuum suggests that the same mechanism dominates the excitation 
of both ionic and PAH bands.

\subsection{NGC\,1808 PAH Profiles Analysis}

\citet{peeters02} and \citet{diedenhoven04} demonstrated that the PAH bands 
in a sample of reflection nebulae, H{\sc ii} regions, young star objects, evolved stars and galaxies
can show different profiles, which they proceeded to classify based on their shape and peak position into three classes. The $A$ class has the peak shifted to the blue, while $B$ and $C$ would show the peak centred and shifted to the red, respectively.
According to these authors, H{\sc ii} regions, non-isolated Herbig AeBe stars and 
young stellar objects generally can be classified as $A$ class for all $3-12\mu$m
PAH emission features. Their galaxy sample shows the same profile as the H{\sc ii}
regions ($A$ class) except for the 11.3$\mu$m profile, which resembles that of 
some evolved stars ($B$ class). They also point out that most planetary nebulae 
and post-asymptotic giant branch stars belong to the second group ($B$ class) 
and just two peculiar post-asymptotic giant branch stars form a third class \citep[$C$ class][]{peeters02,diedenhoven04}.
However, \citet{peeters02} and \citet{diedenhoven04} do not analyse the PAH line profiles 
in the AGNs environment.
These authors have been attributed the PAH different profiles to differences in the physical conditions, composition of the PAH family and/or the accumulated effect of processing in the regions where the emission originates \citep[see also][]{genzel98,diedenhoven04,lebouteiller07,tielens08,sales10}. 
In addition, other authors propose that the trends in the PAH band line profiles and intensities are a function of the PAH molecules size, charge, geometry, and heterogeneity \citep{hony01,bauschlicher08,bauschlicher09}.

\begin{figure*}
\includegraphics[scale=0.7,angle=-90]{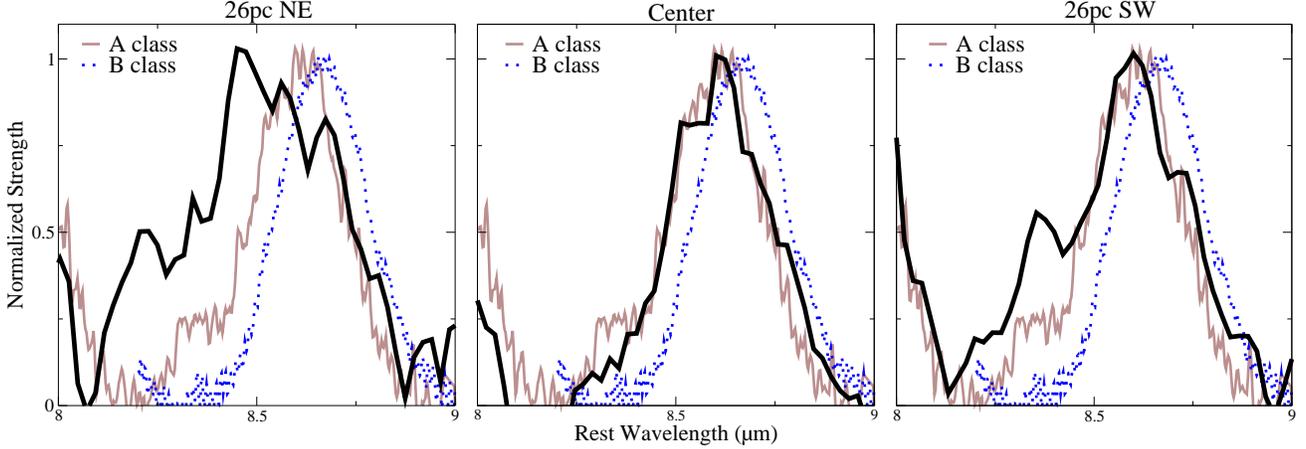}
\caption{Comparison between the NGC\,1808 8.6$\mu$m PAH profiles and the A$_{8.6}$, 
and B$_{8.6}$ Peeters's PAH classes. All profiles have been normalized to the peak 
intensity. The solid black line correspond to the observed PAH profile.}
\label{8profile}
\end{figure*}
\begin{figure*}
\centering
\includegraphics[scale=0.7,angle=-90]{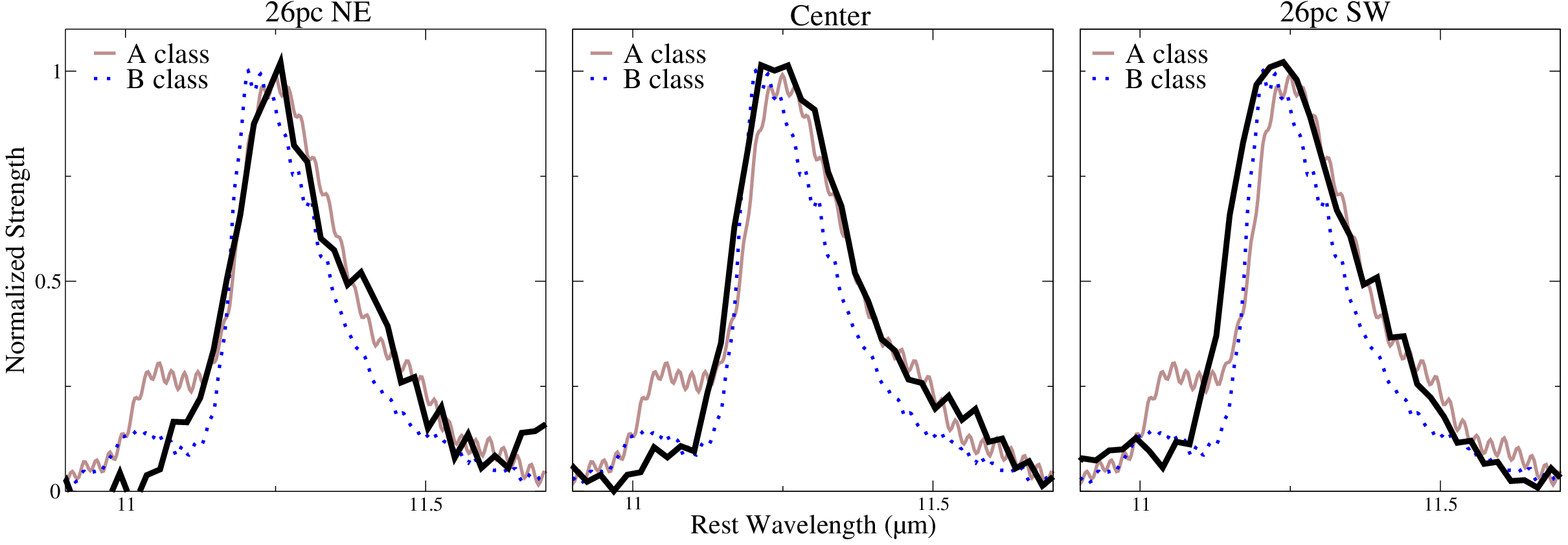}
\caption{Comparison between the NGC\,1808 11.3$\mu$m PAH profiles and the A$_{11.3}$, 
and B$_{11.3}$ Peeters's PAH classes. All profiles have been normalized to the 
peak intensity.}
\label{11profile}
\end{figure*}

In order to investigate the PAH profiles using high spatial resolution ($\sim$26\,pc) of
the Seyfert NGC\,1808, we isolate the PAH emission subtracting a general and local continuum under the 8.6\,$\mu$m and 11.3\,$\mu$m bands,
using the same procedures of \citet{peeters02} and \citet{diedenhoven04}.
Figures \ref{8profile} and \ref{11profile} show the 8.6\,$\mu$m and 11.3\,$\mu$m PAH profiles from our data, compared to 
representative PAH profiles of the $A$ and $B$ \citet{peeters02} classes\footnote{The \citet{peeters02} profiles are
available at http://www.astro.uwo.ca/~epeeters/research-data.shtml.}.

In general, the 8.6$\mu$m and 11.3$\mu$m PAH emission line profiles in NGC\,1808 resemble Peeters' $A$ 
class. However, the 26\,pc SW and 26\,pc NE 8.6$\mu$m PAH profiles are wider
than the  representative $A$ class example, and also shifted to shorter wavelengths. 
Comparison with the theoretical models from \citet{bauschlicher09} would suggest that the
excess emission in the blue wing of the 8.6$\mu$m line profile is evidence that the 
off-nuclear regions at 26\,pc SW and 26\,pc NE contain more ionised PAH material than the 
nuclear region in this galaxy. The same conclusion can be derived 
from the equivalent width of the 8.6$\mu$m assuming that this line is emitted by ionized PAH molecules
\citep[see][]{allamandola99}.


Fig. \ref{11profile} shows that the 11.3$\mu$m PAH profile from the 26\,pc SW spectrum is broader towards shorter wavelengths than the representative $A$ class. According to \citet{bauschlicher08,bauschlicher09}, the emission creating the shorter wavelengths of the 11.3$\mu$m profile would be produced mainly by neutral PAH molecules, and we would conclude that the amount of neutral emitting PAH material in the 26\,pc SW region is larger than in the NGC\,1808 centre. The same conclusion can be derived 
from the equivalent width of the 11.3$\mu$m assuming that this line is emitted by neutral PAH molecules
\citep[see][]{allamandola99}.


In addition, we can see from the values in the Table~\ref{lines1} that the equivalent width (EW)
of the PAH at 8.6$\mu$m, which is due to the ionized PAHs \citep[e.g.][]{puget89,allamandola89,allamandola99,peeters02,tielens08}, 
is larger at 26\,pc SW extraction, while the largest EW of the neutral PAHs 
(see references above)
is at 26\,pc NE extraction. In summary, the comparison of the observed PAH profiles 
and their EWs with the existing observational and theoretical studies indicate
that at 26\,pc SW of the nucleus in NGC\,1808, the ratio of ionised to neutral PAH molecules 
is larger (1.3) than in the nuclear region (0.96). On the other hand, the EW ratio of ionised to neutral PAH
is smallest, EW11.3$\mu$m/EW8.6$\mu$m$\sim$0.7, at 26\,pc NE suggesting that this region 
has more PAHs in the neutral ionization stage.
Interestingly, the region hosting more ionised PAH coincides with the $S1$ X-ray source of \citet{jimenez05},  while the side hosting more neutral molecules  coincides with the extended diffuse X-ray emission
(see Fig.~\ref{fenda}). Overall, the asymmetry in the PAH profiles correlates with regions associated with soft X-ray emission (E\,$<1.5$keV).

Our data make it quite evident that the 8.6$\mu$m and 11.3$\mu$m 
PAH emission (from ionised and neutral molecules) are present in regions of soft and hard X-ray emission,
as well as near to the active nucleus itself (r $\lesssim$30pc). This immediately implies that both ionised and neutral PAH molecules are surviving in the hard X-ray radiation field of the AGN,  as predicted if a dusty structure,
such as the putative torus in the AGN unified model, is present to partially block the strong radiation
field from the central black hole. In this context, it is an interesting issue to investigate if the nuclear spectrum of NGC\,1808 present evidence of  dusty torus that would explain the presence of PAH molecules near to the active nucleus.

\subsection{Is there a Dusty Torus Present in the Central Region of NGC\,1808?}

The presence of a nuclear dusty toroidal structure, mainly composed by silicate and graphite grains,
leaves unmistakable signatures in the observed SEDs of AGN. While the sublimation
of the graphite grains creates IR emission at $\lambda\ge1\mu$m,  the $\sim$9.7$\mu$m
feature observed in emission/absorption is attributed to silicate grains \citep[e.g.][]{barvainis87,pier92,granato94,siebenmorgen05,fritz06,ardila06,riffel09}. Some authors describe the torus as a continuous density distribution \citep[e.g.][]{pier92,granato97,siebenmorgen04,fritz06}, but it has been postulated that, for dust grains to survive in the torus environment, they should be formed in clumpy structures \citep{krolik88} and once provide a natural attenuation 
of the silicate feature \citep[e.g.][]{nenkova02,nenkova08a,nenkova08b,honig06}.

Following \citet{sales11}, we removed the underlying host galaxy contribution from the NGC\,1808 nuclear spectrum by 
subtracting the average spectrum of the two adjacent extractions (apertures -1 and 1 in Fig.~\ref{fenda}).
Moreover, we masked the emission lines and the telluric band region (Fig.\ref{ngc1808}) 
using a simple interpolation, and then compared the resulting pure nuclear spectrum 
against the Nenkova's models \citep[e.g.][]{nenkova02,nenkova08a,nenkova08b}.

These models\footnote{The models are available from http://www.pa.uky.edu/clumpy/} 
assume that the torus is formed by dusty clumps constrained by the following parameters:
(\textit{i}) the number of clouds, \textit{$N_{0}$}, in the torus equatorial radius; 
(\textit{ii}) the optical depth of each cloud in the V band, $\tau_{V}$; 
(\textit{iii}) the radial extension of the clumpy distribution, $Y=R_0/R_d$, 
where $R_0$ and $R_d$ are the outer and inner radius of the torus, respectively; 
(\textit{iv}) the radial distribution of clouds as described by a power law $\propto r^{-q}$;
(\textit{v}) the torus angular width, constrained by a Gaussian angular distribution described by a half-width $\sigma$; 
(\textit{vi}) the observer's viewing angle $i$.

\begin{figure}
\centering
\includegraphics[width=9cm]{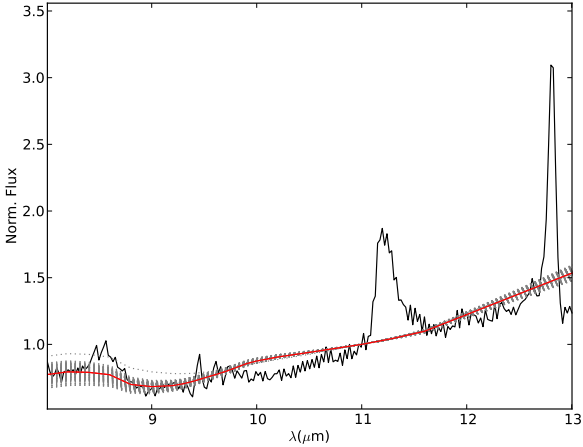}
\caption{Comparison between the NGC\,1808 nuclear spectrum after removal of the 
underlying galaxy contribution (solid black line), and the best fit torus model 
(red line) as defined in the text, normalised at 11.0$\mu$m. The grey dotted lines
correspond to the 10\% of best-fitting SEDs.}
\label{nenkova}
\end{figure}

The ionic and PAH emission lines, as well as the telluric band (Fig.\ref{ngc1808}) 
were ignored for the fit. The final spectrum was compared with the $\rm \sim 10^6$ {\sc clumpy} theoretical SEDs, and the best fit is obtained by searching  for the minimum in the equation:

\begin{equation}
\chi^{2} = \frac{1}{N}\,\sum_{i=1}^{N}\,\left(\frac{F_{obs,\,\lambda_i} - F_{mod,\,\lambda_i}}{\delta_{\lambda_i}}\right)^{2},
\end{equation}
where N is the number of data points in the spectrum, F$_{obs,\,\lambda_i}$ and F$_{mod,\,\lambda_i}$ 
are the observed and theoretical fluxes at each wavelength, and $\delta_{\lambda_i}$
are the uncertainties in F$_{obs,\,\lambda_i}$, calculated as  10\% of $F_{\lambda}$ \citep[see][]{radomski02,mason06,sales11}. 
Both F$_{obs,\,\lambda_i}$ and F$_{mod,\,\lambda_i}$ were normalised to unit at 
11.0$\mu$m, with the uncertainties correctly propagated. Table~\ref{statistic}
show the parameters of the best fit, and Fig.~\ref{nenkova} shows the resulting theoretical SED superimposed on the NGC\,1808 pure nuclear spectrum.

The mean and standard deviation of the model parameters were calculated from the subset of models 
with a $\chi^2$ value within 10\% of the best-fitting result (144 in total). The 
locus of the corresponding theoretical SEDs is plotted as a grey region in Figure~\ref{nenkova}. 
We used a similar  approach  to that of \citet{nikutta09}, who test
5\%, 10\%, 15\% and 20\% $\chi^2$ deviation fractions, and found that 
our methodology results in a narrower locus of acceptable solutions. The 
mean values for the model parameters within the adopted $\chi^2$ 
deviations are shown in Table~\ref{statistic}.

The parameters obtained for the best fit model suggest that NGC\,1808 hosts a dusty 
toroidal structure, where the dusty clouds, with individual optical depth of 
$\tau_{V}$ = 100\,mag, occupy a toroidal volume of $R_0/R_d$ = 10. 
The density profile distribution of the clouds follows a power-law r$^{-1.5}$, 
and there are 15 clouds in the equatorial radius of this torus.
The toroidal structure is constrained in angular width by a Gaussian distribution 
of width $\sigma$ = 70$^{\circ}$.  

\begin{table*}
\centering
\scriptsize
\renewcommand{\tabcolsep}{1mm}
\caption{Resulting Parameters for the fit of the CLUMPY models to the nuclear spectrum of NGC\,1808.}
\label{statistic}
\begin{tabular}{lccccc}
\hline\hline
\noalign{\smallskip}
\noalign{\smallskip}
Parameter & Best Fit & Average & Average & Average & Average\\
& & 5\% & 10\% & 15\% & 20\%\\
\hline
Torus angular width ($\sigma$) & 	70$^{\circ}$ & 	65$^{\circ}$\,$\pm$\,5$^{\circ}$ & 62$^{\circ}$\,$\pm$\,7$^{\circ}$ &61$^{\circ}$\,$\pm$\,8$^{\circ}$ &61$^{\circ}$\,$\pm$\,8$^{\circ}$\\
Radial extent of the torus ($Y$) & 10& 10\,$\pm$\,0	& 11\,$\pm$\,7& 11\,$\pm$\,8&11\,$\pm$\,10\\
Number of clouds in the torus equatorial radius ($N_0$) & 15& 14\,$\pm$\,1	& 13\,$\pm$\,1&13\,$\pm$\,2&12\,$\pm$\,2\\
Power-law index of the radial density profile ($q$) & 	1.5&	1.5\,$\pm$\,0.1 &1.1\,$\pm$\,0.3&0.7\,$\pm$\,0.5&0.6\,$\pm$\,0.5\\
Observers' viewing angle ($i$) & 	90$^{\circ}$& 80$^{\circ}$\,$\pm$\,9$^{\circ}$ &	76$^{\circ}$\,$\pm$\,11$^{\circ}$&74$^{\circ}$\,$\pm$\,13$^{\circ}$& 71$^{\circ}$\,$\pm$\,15$^{\circ}$\\
Optical depth for individual cloud ($\tau_V$) & 	100\,mag & 100\,$\pm$\,0\,mag	& 99\,$\pm$\,6\,mag&99\,$\pm$\,6\,mag&100\,$\pm$\,7\,mag\\
Total number of solution & 	 & 25	& 144& 343& 525\\
\noalign{\smallskip}
\hline
\end{tabular}
\end{table*}

According to our best-fit model,  we would be looking along the torus equator ($i$ = 90$^{\circ}$), resulting in a line of sight  column density of N$_{H}\,\approx\,1.5\,\times\,10^{24}\,cm^{-2}$ and torus outer radius is R$_{0}\sim$0.55\,pc.
We can also estimate the bolometric luminosity from clumpy torus model 
as L$_{bol}$ = 1.9\,$ \times 10^{42}\,$erg s$^{-1}$ \citep[see][]{nenkova08b,sales11}.
This value is a factor of $\sim$6 larger than that inferred 
from X-ray observation \citep[L$_{X-ray}$ = 1.6\,$\times\,10^{40}\,$erg s$^{-1}$][]{jimenez05}
after apply the correction factor of 20 suggested by \citet{elvis94}.
\citet{ramos09} have also found similar results from their clumpy torus model fitting 
(L$_{X-ray}^{bol}$ = 6.8\,$\times\,10^{41}\,$erg s$^{-1}$) of the NGC1808 photometric data.
However, such discrepancy could be attributed to the attenuation from circumnuclear dust \citep[e.g.][]{krabbe94}
included in the larger (850\,pc) aperture used for the X-ray (850\,pc) used by \citet{jimenez05}.


Note that the high column density derived from the models would be enough to block the hard radiation from the central source \citep{avoit92,bvoit92}, explaining how the observed PAH molecules can survive so close (r$\lesssim$26\,pc) from the 
Seyfert nucleus.

\section{Summary and Conclusions}\label{conclusions}

In this paper we analyse  N band spectra of the Seyfert 2 NGC\,1808, obtained with the Gemini T-ReCS  
at a spatial resolution $\sim$26\,pc. The central regions of this galaxy show extended PAH band emission, and 
conspicuous differences in the PAH line profiles are detected in scales of dozens of parsecs. Our main conclusions are:

\begin{enumerate}
  \item The spectra of the unresolved nucleus and of regions located 26\,pc in both NE and SW directions along the slit show
  intense 8.6$\mu$m, 11.3$\mu$m and 12.7$\mu$m PAH emission bands. The spectra also show 
  [Ne{\sc\,ii]}12.8$\mu$m and H$_{2}\,S$\,(2)12.27$\mu$m emission lines. Both ionic and 
  molecular emission are extended up to 70\,pc from the nucleus.
  
  \item The PAH, [Ne{\sc\,ii]}12.8$\mu$m, and 10$\mu$m continuum emission decrease radially in intensity. The similarity in their luminosity profiles would indicate that their excitation mechanism is the same.

    \item The analysis of the PAH emission lines shows that the 8.6$\mu$m and 11.3$\mu$m line profiles in the central region are similar to Peeters's $A$ class. However, the profile of the lines originating from the  26\,pc SW and 26\,pc NE regions are different, implying that in the ratio of ionised to neutral PAH molecules in the 26\,pc NE region is higher than in the 26\,pc SW one.

   \item Strong PAH molecular emission is present in the nuclear extraction, as well as in the off-nuclear regions coincident with hard and soft X-ray emission. This suggests that the PAH molecules in the nucleus of NGC\,1808 are able to survive the AGN hard radiation field.
   
  \item The nuclear N band spectrum after subtraction of the underlying galaxy contribution can be represented by a clumpy torus model \citep{nenkova08b}. The torus in NGC\,1808 would present a  toroidal volume of 
  $R_0/R_d$ = 10, with each cloud having an optical depth $\tau_V$ = 100\,mag and the 
  distribution of the clouds following a $r^{-1.5}$ power-law. 
    
  \item According to the best-fit torus model, our line of sight would be along the torus equatorial radius ($i = 90^{\circ}$), resulting a hydrogen column density of N$_{H}\,\approx\,1.5\,\times\,10^{24}\,cm^{-2}$. The torus would have an outer radius of R$_{0}\sim$0.55\,pc. Our modelling is consistent with NGC\,1808 hosting a Sy~2 nucleus.

\end{enumerate}

\section*{Acknowledgements}
\label{acknowledgements}

The authors thank for the anonymous referee for many useful suggestions.
MP would like to acknowledge the support from CNPq (grant 308985/2009-5). RR acknowledges 
funding from FAPERGs (ARD 11/1758-5) and CNPq. Based on observations obtained at the Gemini 
Observatory, which is operated by the Association of Universities for Research in Astronomy, 
Inc., under a cooperative agreement with the NSF on behalf of the Gemini partnership: the 
National Science Foundation (United States), the Science and Technology Facilities Council 
(United Kingdom), the National Research Council (Canada), CONICYT (Chile), the Australian 
Research Council (Australia), Minist\'{e}rio da Ci\^{e}ncia e Tecnologia (Brazil), and 
Ministerio de Ciencia, Tecnolog\'{i}a e Innovaci\'{o}n Productiva (Argentina).
We thank Charles Bonatto and Isabel Aleman for helpful discussion.

\end{document}